\documentclass[12pt,preprint]{aastex}
\usepackage{graphicx}
\usepackage{amssymb}
\begin{document}

\title{The Broad-Band Spectrum and Infrared Variability of the Magnetar
  AXP 1E~1048.1$-$5937}
\author{Martin Durant and Marten H. van Kerkwijk}
\affil{Department of Astronomy and Astrophysics, University of
  Toronto\\  60 St. George St, Toronto, Canada}

\begin{abstract}
We present photometry of the Anomalous X-ray pulsar 1E~1048.1$-$5937
in the infrared and optical, taken at Magellan and the VLT. The object
is detected in the I, J and K$_\textrm{s}$ bands under excellent
conditions.  We find that the source has varied greatly in its
infrared brightness and present these new magnitudes. No correlation
is found between the infrared flux and spin-down rate, but the
infrared flux and X-ray flux may be anti-correlated.  Assuming nominal
reddening values, the resultant spectral energy distribution is found
to be inconsistent with the only other AXP SED available (for
4U~0142+61). We consider the effect of the uncertainty in the
reddening to the source on its SED.  We find that although both
the X-ray and infrared fluxes have varied greatly for this source, the
most recent flux ratio is remarkably consistent with what is is found
for other AXPs.  Finally, we discuss the implications of our findings
in the context of the magnetar model.
\end{abstract}
\maketitle

\section{Introduction}
Over the last decade or so, evidence has been mounting for the existence
of a class of neutron stars called the {\em magnetars} (Thompson \& Duncan,
1996).
They have enormous external magnetic fields of order $\sim10^{14}$G
(and even larger internal field), the decay of which powers luminous high
energy radiation. Examples of
this type of astrophysical source are the Soft Gamma-ray Repeaters
(SGRs), which give sporadic bursts of hard X-ray/soft gamma rays as
well as rare, very luminous ($\sim 10^{44}$erg) ``giant flares''; and
the Anomalous X-ray Pulsars (AXPs), so called because their luminosity far
exceeds the spin-down value, and no binary companion is seen. This
implies that the source of power cannot be rotational energy and that
accretion is excluded. 
Both the SGRs and AXPs
have spin periods of order $P=10$s and derivatives
$\dot{P}=-10^{-10}\ldots -10^{-12}$, and are inferred to be young
from their energetics and spin-down, and from supernova remnant
associations in some cases. The two groups have been linked by the
discovery of persistent emission from SGRs that is similar to the AXPs
and bursting behaviour in
the AXPs. See Woods \& Thompson (2004) for a review.

It is with the discovery of optical and infrared counterparts (Hulleman et
al., 2000) that models for AXPs other than the magnetar became
untenable. Hulleman et al. (2004)
were the first to produce a spectral energy distribution for the AXP
4U~0142+61, showing an intriguing hint of a spectral feature: a sharp
break in the optical.
Furthermore, Kern \& Martin (2002) discovered, optical pulsations with
pulsed fraction of the order 25\%,
modulated at the X-ray period. 

Although in the magnetar model all the details are not yet
ironed out, it does explain how such emission can arise from cyclotron
emission by ions in the outer magnetosphere (Thompson, 2004,
priv. comm.; Thompson, Lyutikov \& Kulkarni, 2002).
From the magnetar theory, it is not entirely clear whether one would
expect the X-ray and infrared fluxes to be correlated. Such a
relationship has  been seen for the AXP
1E~2259+586, which has recently been shown to have correlated X-ray and
infrared fluxes following an X-ray bursting episode (Tam et al., 2004).
The return to the quiescent flux was found to occur on
the same time-scales in both bands, and the flux ratio to remain
roughly constant throughout the active episode. 

1E~1048.1$-$5937, is a 6.4s AXP in the field of the Carina Nebula, one
of the two AXPs to have shown SGR-like outbursts to date and also the
most noisy in terms of its timing characteristics (Kaspi et al., 2001;
Gavriil, Kaspi \& Woods, 2002). This, as well as its relatively hard
spectrum, make it the most SGR-like of the AXPs (Woods \& Thompson,
2004).  A possible infrared counterpart to 1E~1048.1$-$5937 was
observed by Wang \& Chakrabarty (2002; henceforth WC02), who found the
magnitudes: $J=21.7(3)$, $H=20.8(3)$ and
$K_{S}=19.4(3)$. Intriguingly, the source was not detected by Israel
et al. (2002) down to limits of $J\sim 23$, $H\sim 21.5$ and
$K_{S}\sim 20.7$.  Ergo the object is highly variable, and therefore
of particular interest.

Here we present further deep photometry in both the optical and the
infrared, showing detections in three bands at levels far fainter than
those of WC02, but consistent with Israel et al.'s limits. It may be
that WC02 observed the object in a state similar to 1E~2259+586 in
which its infrared flux was enhanced because of a preceding
outburst. If so, this work probably shows the spectral energy
distribution of the object in its quiescent phase. 

Below, we will first present imaging observations performed at the VLT and
Magellan, and describe their analyses. Next, we describe the implications
of these in three main areas: the variability, spectral
energy distribution, and X-ray to infrared flux ratio. We conclude with a
brief discussion.

\section{Observations and Analysis}
\begin{figure}
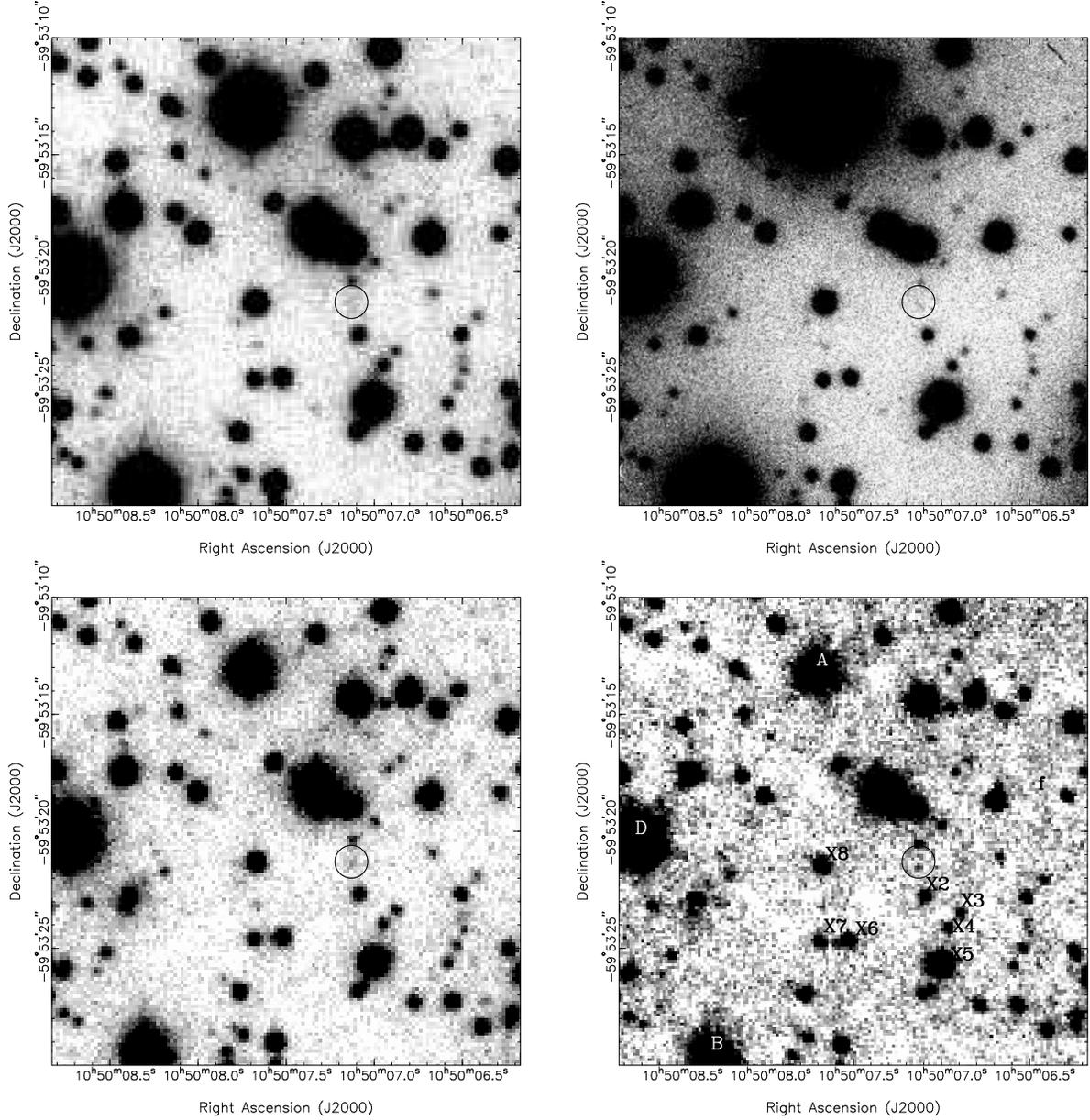

\begin{center}
\parbox{0.49\hsize}{
\includegraphics[width=\hsize,angle=270]{f1a.eps}
\includegraphics[width=\hsize,angle=270]{f1b.eps}}
\parbox{0.49\hsize}{
\includegraphics[width=\hsize,angle=270]{f1c.eps}
\includegraphics[width=\hsize,angle=270]{f1d.eps}}
\caption[0]{Images of the field of 1E~1048.1$-$5937, in all bands with
  detections: J-band (top left),
  I-band (top right) and the K$_\textrm{s}$-band (bottom) centred on the
  position of the AXP. The left-hand
  images were taken with the VLT and the right-hand ones with
  Magellan. The $0.74''$ radius error circle is shown on each
  image. Note that the horizontal band-like structure in the Magellan
  K$_\textrm{s}$ image is due to a irreproducible ``pattern noise''
  that existed on the PANIC instrument at the time (see text).}\label{stars}
\end{center} 
\end{figure}

Infrared imaging (J, H and K$_\textrm{s}$ bands) was performed of the field of AXP
1E~1048.1$-$5937 using ISAAC at the VLT. At Magellan, further imaging data were
taken both in the optical (I, z$^\prime$ and a wide `VR' filter) and infrared
(J and K$_\textrm{s}$) using MagIC
and PANIC, respectively. See Table \ref{obslog} for a list of
observations and the conditions for each, and Figure \ref{stars} for
images of each of the detections.

\begin{deluxetable}{lccccccc}
\tablecaption{Observation log of 1E~1048.1$-$5937. \label{obslog}}
\tablewidth{0pt}
\tabletypesize{\footnotesize}
\tablehead{
\colhead{Date} & \colhead{Telescope,} & \colhead{Mid-exposure} &
  \colhead{Band} & \colhead{Integration} & \colhead{Seeing}\tablenotemark{a} &
  \colhead{Magnitude} & \colhead{$\nu F_{\nu}$} \\ 
 &  \colhead{instrument} & \colhead{(UT)} & & \colhead{time (s)} &
  \colhead{(arcsec)} & & (erg s$^{-1}$ cm$^{-2}$)}
\startdata
24 April 2003 & VLT & 01:40\tablenotemark{b} & K$_\textrm{s}$ & 2772 & 0.34 & 21.3(3) &
  $2.6(9) \times 10^{-15}$\\
 & ISAAC & 01:45\tablenotemark{b} & H & 1848 & 0.47 & $>$21.3 & $< 5.6 \times 10^{-15}$\\
 & & 02:15 & J & 3255 & 0.43 & 23.4(4) & $1.6(8) \times 10^{-15}$ \\
6 June 2003 & Magellan & 03:50 & I & 7200 & 0.33 & 26.2(4) & $3.0(13) \times 10^{-16}$\\
 & MagIC & 05:35 & VR & 2700 & 0.45 & $>$26.0\tablenotemark{c} & $< 6.8 \times 10^{-16}$\\
 & & 06:30 & z$^\prime$  & 2700 & 0.44 & $>$24.2\tablenotemark{d} & $< 3.0 \times 10^{-15}$\\
7 June 2003 & Magellan & 03:30 & K$_\textrm{s}$ & 2300 & 0.38 &
  $>$20.9 & $< 3.8 \times 10^{-15}$\\
 & PANIC & 06:00 & H & 3780 & 0.34 & $>$20.8 & $< 9.0 \times 10^{-15}$\\
 & & 05:00 & K$_\textrm{s}$ & 1375 & 0.30 & 21.5(4) & $2.2(10) \times 10^{-15}$\\
\enddata
\tablecomments{~Limits are at 95\% confidence levels.}
\tablenotetext{a}{Full width at half maximum.}
\tablenotetext{b}{The VLT images were taken alternating between the
  filters, hence the close mid-exposure times.}
\tablenotetext{c}{This is on a Vega-like magnitude scale, with
  $F_\nu(VR=0)=3.33(2) \times 10^{-20}$ erg s$^{-1}$ cm$^{-2}$ Hz$^{-1}$
  (see text).}
\tablenotetext{d}{This magnitude is not on the Vega, but on the
  AB-system (see text).}.
\end{deluxetable}

\subsection{Astrometry}
The images below were referenced to the International Celestial
Reference System through identifying stars in the USNO B1.0 catalogue
on a short (30s) I-band image. Sixty-five stars were
cross-identified, and a solution found with RMS deviations $\sim
0.3''$ in each coordinate (after rejecting objects with residuals greater
than $0.6''$, leaving 49 good matches), and thus an astrometric error of $0.3/
\sqrt{49} = 0.04''$ in
connecting to the USNO reference frame. This in turn has a systematic uncertainty of
$0.2''$ (Monet et al. 2003) in connecting to the ICRS. Thus the total uncertainty in
astrometry is $\sim 0.2''$ in each coordinate.

The uncertainties in connecting the coordinates of the short I-band exposure
to the rest of the frames are negligible in comparison with the above
errors. The $0.6''$ error radius, at 90\% confidence, in the {\em Chandra}
position of 1E~1048.1$-$5937 (Zombeck et al., 1995; WC02) combined in
quadrature with the astrometric error above, corresponds to an
error of $\sim 0.74''$ on any of the images in Figure \ref{stars}.

\subsection{VLT/ISAAC}
On the night of 23 April 2003, infrared images of the field of
1E~1048.1$-$5937 were taken with ISAAC (Infrared Spectrometer And Array
Camera; Moorwood, 1998) on Antu, unit telescope 1 of the Very Large Telescope.
We used the Long Wavelength (LW) arm with the Aladdin
1k$\times$1k InSb infrared array, with pixel size $0.148''$.
Seeing conditions were excellent. See Table \ref{obslog} for a list of
integration times and seeing for each image. The source was detected
in the long J- and K$_\textrm{s}$-, but not in the shorter H-band
observations.

Standard reduction was carried out in order to subtract the dark current, flat
field (using median averages derived from the science frames) and
combine the frames using {\tt IRAF}. Photometry was
performed using {\tt DAOPHOT~II} (Stetson, 1987).

In order to calibrate the frames, aperture photometry was performed on
short-exposure images of three standard stars obtained on the same
night. The aperture size was set large enough to include the majority
of the flux from the star, as the sky noise was negligible in
comparison; this established the magnitude zero point for each of the
frames. The airmass of these exposures was in every case within 0.1 of
the science exposures, so an atmospheric extinction correction was not necessary.
Colour terms were negligible.  In order to find the
magnitude of the stars in the science field, the aperture correction
was calculated using the difference between the
magnitudes produced by the {\tt allstar} task and those derived using
the same large aperture used for the standards for several bright,
relatively isolated stars in the frame (after the removal of fainter
neighbours). See Table \ref{phot} for a list of the
magnitudes of field stars around the AXP, with the numbering scheme as
in WC02.

The errors on measurements, and also the limiting magnitudes where the
source was not detected, were determined empirically. A routine was
run which inserted stars of known magnitude (scaled from the PSF) into
areas of the image free of stars, and then the new image was passed
through the same analysis procedure as for the science frame and the
standard deviation of the measured magnitudes was calculated for the
inserted stars.  The latter is used as the measurement error for that known
magnitude. It incorporates the fact that fainter stars will be
spread over a wider magnitude range as measured, and that near the
magnitude limit the faintest stars will not be measured at all. The
method gets around any assumptions of the flat field accuracy or
profile errors, which are assumed constant parameters in the error
model used by {\tt DAOPHOT}. Whilst such parameters could be
fitted for each image in turn, and then used to determine the
errors, we found that in every case the function of error versus
magnitude found by {\tt DAOPHOT} deviated both from our standard deviations and from photon
noise whatever the choice of flat-field and profile error. This turned
out to be significant at the very bright end and near the magnitude
limit. 

We find, in the manner described above, the following magnitudes for
the star X1, the AXP counterpart: $H>21.3$, $J=23.4(4)$,
$K_S=21.3(3)$. The limit in H corresponds to 95\% confidence.

\subsection{Magellan/PANIC}
On the night of 6 June 2003, PANIC (Persson's Auxilliary Nasmyth
Infrared Camera; Martini et al., 2004), the 1k$\times$1k infrared imaging
array with $0.125''$ pixels on the Magellan Clay Telescope
\footnote{see {\tt
http://www.ociw.edu/lco/magellan/instruments/PANIC/panic/}}, 
was used to acquire images of the field of 1E~1048.1$-$5937 in the H- and
K$_\textrm{s}$-bands under good conditions.  Two separate imaging runs were performed in the
K$_\textrm{s}$-band through the night, resulting in two images for
analysis. The AXP
was detected in only one of the K$_\textrm{s}$ frames and not at all
in H; see table \ref{obslog}. 

There was a roughly periodic pattern noise in the PANIC detector at
this time. This noise was not reproducible, and its effect can be seen
in our final processed images in Figure \ref{stars} as a band-like
variation in the brightness of the sky.

The images were processed and analysed as above, unsing three standard
stars. We find the following magnitudes for the AXP: $K_S=21.5(4)$,
$K_S>20.9$, and $H>20.8$.

\subsection{Magellan/MagIC}
On 5 June 2003, MagIC (The Raymond and Beverly Sackler Magellan
Imaging Camera\footnote{see {\tt
    http://occult.mit.edu/instrumentation/magic/}}; Shectman \& Johns, 2003), 
a 2k$\times$2k CCD with quad readout
amplifiers and $0.069''$ pixels  on the Clay
Telescope, Las Campanas was used to image 1E~1048.1$-$5937 in the I-
and z$^\prime$-bands. A set of images were also taken using the 
custom `VR' filter installed on MagIC (a wide, roughly rectangular pass band
covering much of V and R; see Jewitt, Luu \& Chen, 1996). For the I
band, about 90\% of the integration time was at excellent
seeing conditions of $< 0.4''$, whereas the average seeing was
$0.44''$ in VR and $0.46''$ in z$^\prime$; see Table
\ref{obslog}. The I-band images with the brightest sky and worst
seeing were excluded from the stacking process, leaving a total of 
integration time of 7200s.

Again, the frames were processed and combined using {\tt IRAF-DAOPHOT
II}, with bias subtraction and trimming performed by the {\tt
ccdmagic} task provided by the observatory and flat fields again
derived from the science frames
(this proved more successful than screen flats).
Magnitudes were calibrated using several standard stars and errors calibrated
empirically as before (following the ISAAC experience
above). 

Seven photometric standard stars for this run were taken from the E5
field (Stetson, 2000), at an airmass close to the mean airmass of the
science exposure. The I-band zero point was a simple average over the
values inferred from the well exposed stars on the frame.

The source was not detected in the VR-band. For calibration of this
non-standard filter, the E5 standard field was also imaged
with the VR filter, and also the standard V and R filters to check for
consistency. The VR zero point and effective wavelength were deduced
in the following manner: the standard stars were assumed to have
power-law spectra over their V\ldots R range, so that the VR magnitude
would be a simple interpolation, $VR = (1-b)V + bR$, where the
parameter $b$ is to be found. Next, VR was defined to be such that
the colour-term coefficient is zero: $VR = vr + z_{VR}$
(where $vr$ is the instrumental magnitude and $z_{VR}$ the zero
point). In effect, this makes this
particular VR filter the prototype for this magnitude scale. Finally,
the value of $b$ was found which minimised variance about an average
value for the zero point. This minimum value was found to be $b=0.51(7)$,
yielding the values: $\lambda_{VR}=0.592(7)\mu$m, $z_{VR}=27.9(4)$ (for 1e-/s),
and $F_\nu(VR=0)=3.33(4) \times 10^{-20}$erg s$^{-1}$ cm$^{-2}$ Hz$^{-1}$.

The z$^\prime$ band is on the magnitude AB system, and is based on a
different set of standard stars 
for its basic calibration. In order to find the zero point, well-known
empirical transformations between z$^\prime$ and the VRI bands were
used (Smith et al., 2001) to find the
magnitudes of each of the standard stars. These were compared with the
measured magnitudes of the standard field with the z$^\prime$ filter.

The AXP counterpart was found to have 
magnitudes $I=26.2(4)$, $VR > 26.0$ and $z^\prime > 24.2$.

\begin{deluxetable}{lcccc}
\tablecaption{Photometry for selected stars in the field of
  1E~0148.1$-$5937.\label{phot}}
\tablewidth{0pt}
\tablehead{
\colhead{Star id\tablenotemark{a}} & \colhead{I} &
  \colhead{J} & \colhead{H} & \colhead{K$_\textrm{s}$}\\}
\startdata
X1 & 26.2(4) & 23.4(4) & $>$21.5 & 21.3(3) \\
X2 & 23.09(2)& 20.44(5)& 19.51(5)& 19.21(4)\\
X3 & 24.85(14) & 21.7(2) & 20.64(15) & 20.00(8)\\
X4 & 23.54(4)& 21.0(1) & 20.12(9)& 19.80(6)\\
X5 & 18.51(2)& 16.83(2)& 16.34(3)& 16.28(2)\\
X6 & 22.24(2)& 19.66(4)& 18.67(4)& 18.46(2)\\
X7 & 22.59(3)& 20.21(5)& 19.35(5)& 19.11(3)\\
X8 & 20.61(2)& 18.63(2)& 18.05(3)& 17.95(2)\\
f & 22.95(3) & 20.46(5)& 19.61(5)& 19.24(3)\\
A & 15.59(2) & 14.68(2)& 14.41(3)& 14.46(2)\\
B & 16.47(2) & 14.97(2)& 14.65(3)& 14.59(2)\\
C\tablenotemark{b} & 17.42(2) & 15.96(2)& 15.52(3)& 15.54(2)\\
D\tablenotemark{b} & 17.11(2) & 14.57(2)& 13.69(3)& 13.46(2)\\
\enddata
\tablecomments{I-band magnitudes are from the Magellan data, the rest
  from the VLT.}
\tablenotetext{a}{As labelled in WC02. Figure \ref{stars} shows the
  locations of some of these sources within the field.}
\tablenotetext{b}{There appears to have been an error in WC02, where
  the IR magnitudes for Stars C and D were switched (Wang, Z., 2005,
  priv. comm.)}
\end{deluxetable}

\section{Results and Implications}

\begin{figure}
\begin{center}
\parbox{0.95\hsize}{
\includegraphics[width=\hsize]{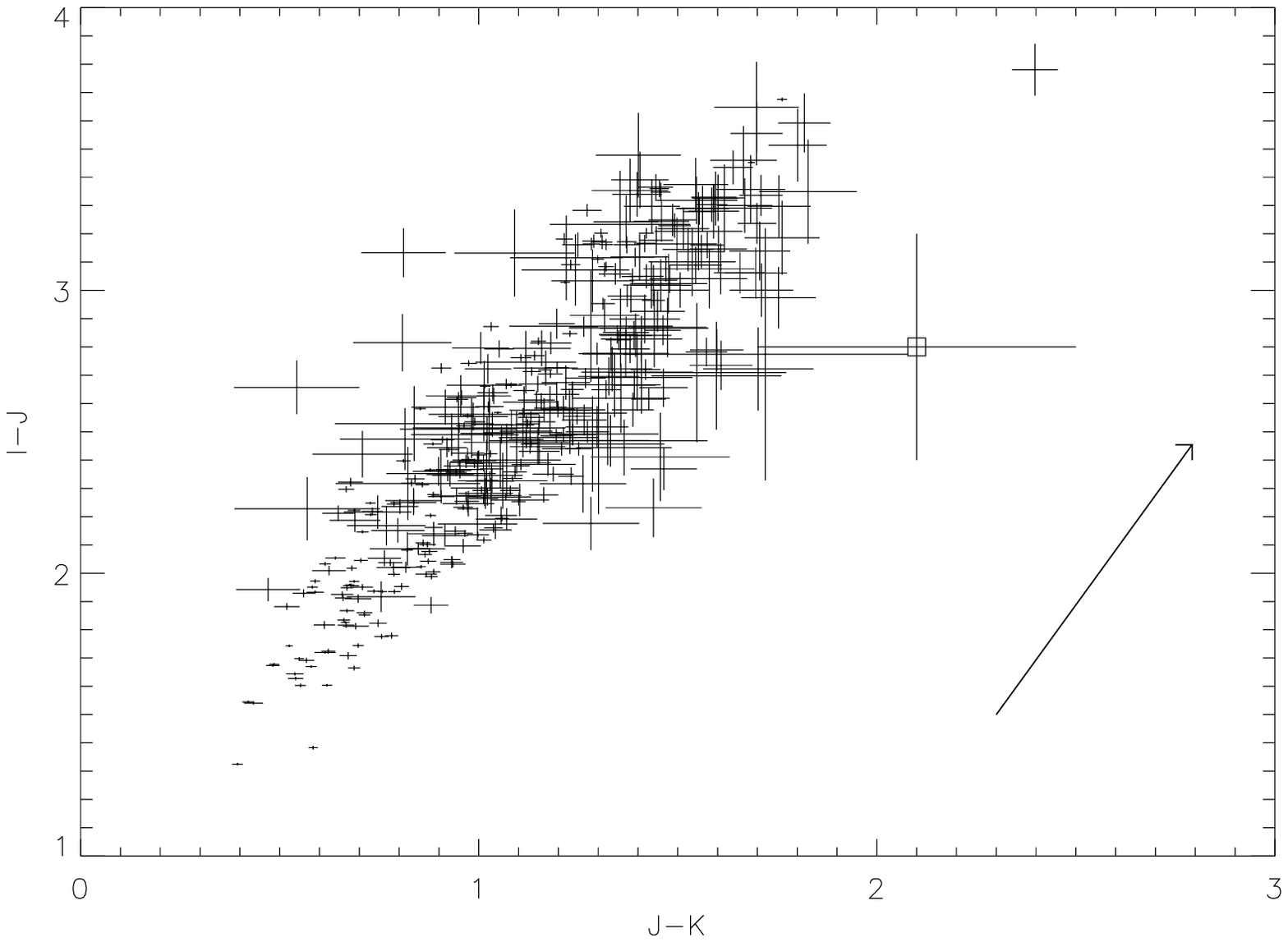}
\caption{Colour-colour plot of stars in the field of
  1E~1048.1$-$5937. The AXP (star X1) is labelled with an open square
  and the arrow shows the direction of increasing reddening. The
  outlier in the top-right is probably a genuine, highly reddened
  background star and a 3$\sigma$ deviation; it lies far from the
  error circle.}\label{c-c}}
\end{center} 
\end{figure}

Figure \ref{c-c} shows a colour-colour diagram for stars in the field
of 1E~1048.1$-$5937. The AXP is clearly offset from the bulk of the stars.
Its J-K$_\textrm{s}=2.1(5)$ colour is similar to that found by WC02
(who found J-K$=2.3(4)$), despite the
fact that it is $\sim2$mag fainter, and to that of the brightest AXP, 4U~0142+61.
Given the variability and peculiar colours, we believe there is no
longer any doubt that the source, labelled X1 by WC02, lying at the
centre of the error circle derived from {\em Chandra} is indeed the
infrared/optical counterpart to the AXP. 

We now discuss the inferred spectral energy distribution, X-ray to
infrared flux ratio and variability. The spectral
energy distribution and X-ray to infrared flux ratio of the AXP depend
on our assumption for the amount of reddening to the source. Our
knowledge on this matter is discussed in Section \ref{red}.

\begin{figure}
\begin{center}
\parbox{0.85\hsize}{
\includegraphics[width=\hsize]{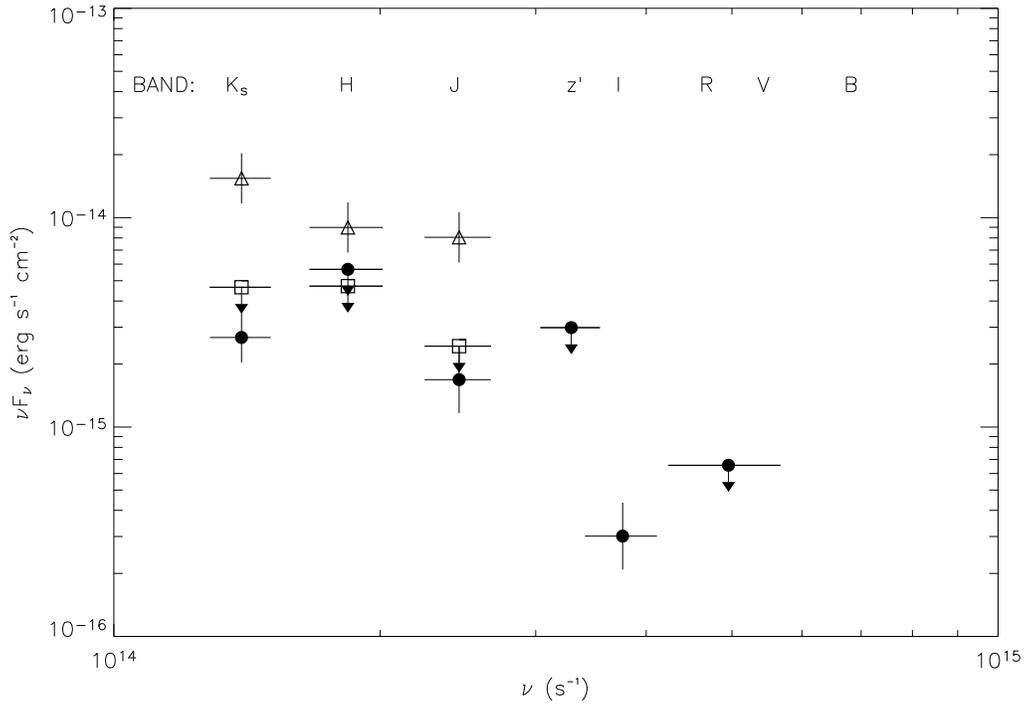}
\caption{Photometry for 1E~0148.1$-$5937,
  as measured. The filled circles represent this work (where the H-,
  z$^\prime$- and VR-band
  points are the respective best limits, and for K$_\textrm{s}$
  only the more
  accurate measurement, from the VLT, is shown); open triangles are
  those of WC02 and open squares the limits established by Israel et
  al. (2002).}\label{sed}}
\end{center}
\end{figure}

\subsection{Variability}\label{varysec}
As has already been noted by Israel et al.
(2002), 1E~1048.1$-$5937 has shown large variability in its infrared
flux. The K$_\textrm{s}$-band magnitudes presented here are much
fainter than those given in WC02, but consistent with the limits
found by Israel et al. This is illustrated in Figure \ref{sed}. 

\begin{figure}
\begin{center}
\parbox{\hsize}{
\includegraphics[width=\hsize]{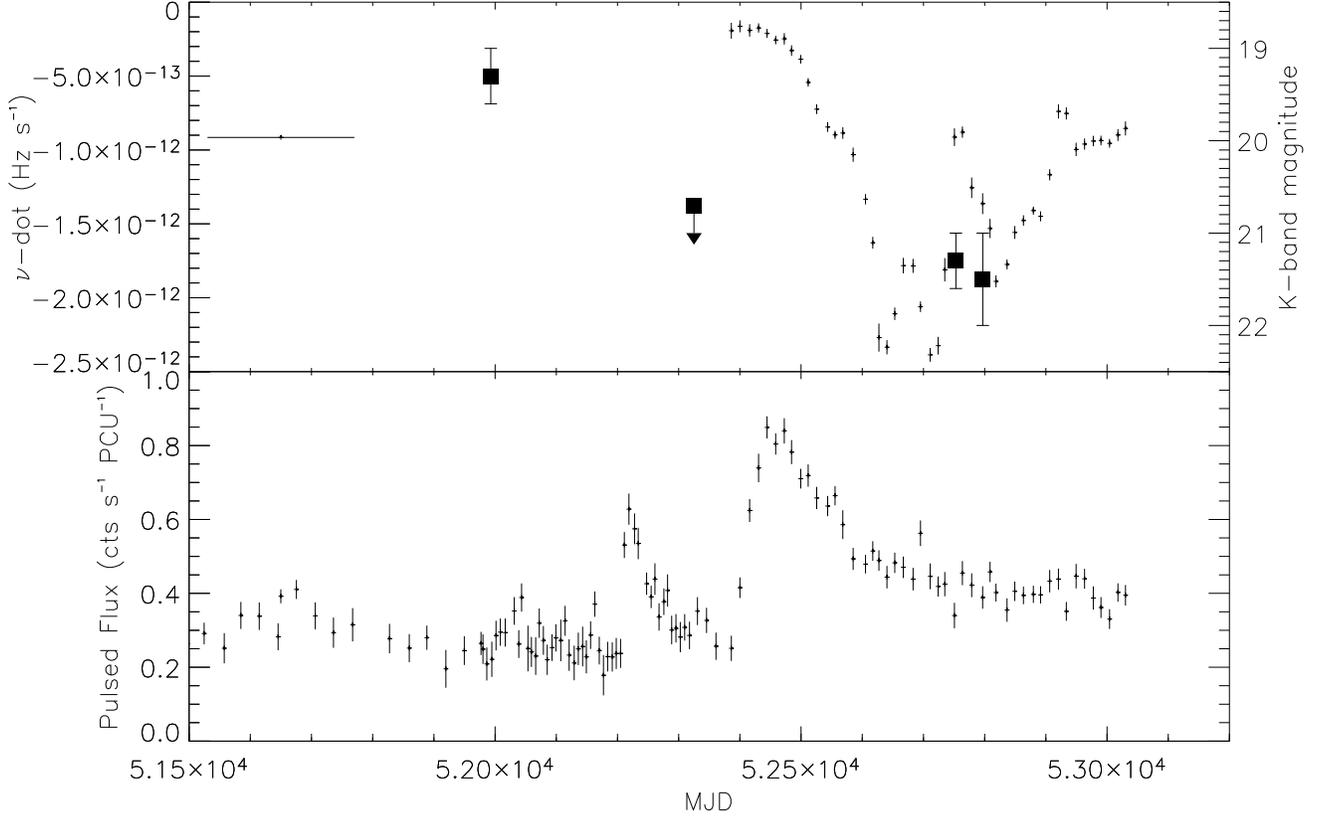}
\caption{Spin-down rate, $\dot{\nu}$ (top panel, crosses),
  K$_\textrm{s}$-band apparent magnitude (top panel, filled 
  squares) and pulsed flux (lower panel) as a function of time. The spin-down and
  flux data are taken from Gavriil \& Kaspi (2004). The wide point
  centred at $(MJD=51650 , \dot{\nu}=-9.15\times10^{-13})$ represents a
  period in 2000 throughout which phase-connected timing was possible
   (Kaspi et al., 2001). At MJD$\sim$52400, the observing strategy was
  changed to enable spin-down measurements without long time-baseline
  phase coherence.}\label{vary}}
\end{center} 
\end{figure}

1E~1048.1$-$5937 is currently part of a regular monitoring programme using the
Rossi X-ray Timing Explorer (RXTE), and has been
observed many times over the last two years, at particularly closely
spaced intervals during the time of the observations above
(Gavriil \& Kaspi, 2004). Marked changes are seen during this
period in pulsed flux\footnote{Due
to nearby strong X-ray sources
and the non-imaging nature of the RXTE observations, only the pulsed flux
could be measured} and spin-down rate ($\dot{\nu}$). The hardness ratio
shows no clear variations during this time. Figure
\ref{vary} shows the pulsed flux and spin-down rate, with K-band
photometry points as a comparison. 
One sees a hint of anti-correlation between the
X-ray and infrared fluxes, and no obvious relationship between the
infrared flux and spin-down rate. As Gavriil \& Kaspi (2004) stated,
X-ray flux and spin-down rate are clearly not related.

For another source also in the RXTE monitoring programme, 1E~2259+586,
recent results indicate that X-ray and infrared fluxes {\em were}
correlated in the period following multiple X-ray bursts and increased
activity (Tam et al., 2004) with both spectral bands showing
increased persistent fluxes, which decayed towards the pre-burst values on the
same time-scale. This suggests that the mechanism at work in
1E~2259+586 is different from that of 1E~1048.1$-$5937, possibly because
of the bursting activity (note that
bursting activity cannot be excluded by the RXTE observations of
1E~1048.1$-$5937).

\subsection{Distance and Reddening}\label{red}
The intrinsic spectrum that we derive for the AXPs depends strongly on the value that we
assume for the interstellar reddening along the line of sight. Since
the reddening to all of the magnetars in the plane of the Galaxy have
nominal reddening $A_V\geq 5$
(which includes all of the AXPs, apart from the recently discovered
CXOU~J010043.1-721134, Majid et al., 2004), the extinction correction is huge and
has a profound impact on the spectrum derived.
As an example, the inferred
optical spectrum of 4U~0142+61 changes from Raleigh-Jeans like to flat
(in $\nu F_\nu$) for a change $A_V=5$ to $A_V=2$ (Hulleman et al.,
2004). It is therefore
important to consider this as part of the analysis.

Usually, the model for the intrinsic X-ray spectrum
is used to infer the amount of extincting material (given as the
Hydrogen column, $N_H$), and assuming the relationship derived by
Predehl \& Schmitt (1995) to calculate the reddening due to dust.

For 1E~1048.1$-$5937, Tiengo et al. (2002) found 
an apparently featureless X-ray spectrum well-fitted by an
absorbed power law ($\nu F_\nu \sim \nu^{-\alpha}$, $\alpha\sim 2-3$)
plus blackbody ($kT \simeq 0.64$ keV). This model for the X-ray
spectrum yields an extremely high value for the extincting column to
the source of $N_H=1.04(8)\times10^{22}$ cm$^{-2}$, comparable to values of
total extinction seen through many lines of sight through the galactic
plane (but see also below).
In this way, values typically of order $A_V \sim 5.8$ have been arrived
at (WC02) for 1E~1048.1$-$5937. Comparing this estimate to the typical
reddening to stars
in the Carina nebula ($A_V\sim 2$), Seward et al. (1986) argued that the AXP must
lie behind the Carina Nebula, and thus placed a lower limit on its
distance of $d\gtrsim 2.8$ kpc.

There exist other independent methods for estimating the run of
reddening with distance in any given line of sight. For our field, we
found the following relevant measurements.

Neckel, Klare \& Sarcander (1980) measured the average function of
reddening with distance, $A_V(d)$, in many galactic-plane fields
including one containing this source. They find that as a function of
distance, the visual extinction is constant between 1.5kpc -- 6kpc at
$A_V\simeq1$. It should be stressed, however, that this value is an
average over their field (size of order $\sim$square degrees), and
one could argue that measurement biases
would favour a lower value of extinction at any given distance in an
area of highly variable dust content.

Carrero et al. (2004) performed colour-magnitude studies of
three nearby open clusters in the Carina complex (approximately
$1^\circ$ from the region of interest). For
the clusters Trumpler 16 and Trumpler 14, they find that $A_V=$2.0(1) and
1.8(7) and distances of $d=$3.9(5) and 2.5(3) respectively. This is in
a region where the dust density is visibly greater than the
surrounding sky; for 1E~1048.1$-$5937 to have a greater extinction, this
suggests a more constraining lower bound on its distance. It should be
noted however, that the authors of this study stress the variability
of reddening from star to star in the clusters.

Finally, Merehgetti, Caravea \& Bignami (1992) find that one of the brighter
stars in the field is probably very highly redenned ($A_V >
7$), based on two interstellar absorption bands (4428  and
5778\AA) in its spectrum. The distance to this star is unknown.

To summarise, there is reason to question our assumption that the value
of $N_H$ gained through the fitting of the X-ray spectrum gives an
appropriate value of reddening; such a suggestion has also arisen in
the case of the brightest of the AXPs, 4U~0142+61. Hulleman, van
Kerkwijk \& Kulkarni (2004) suggest that values in the range
$A_V=2\ldots6$ are possible. A similar uncertainty was also suggested
in the case of AXP 1E~2259+586 (from the measurements of optical
filaments in its
associated supernova remnant, Fesen \& Hurford, 1995). As we see below,
the value we use to de-redden
flux measurements will alter the shape of the resultant spectrum
drastically.

\subsection{Spectral Energy Distribution}
\begin{figure}
\begin{center}
\parbox{0.85\hsize}{
\includegraphics[width=\hsize]{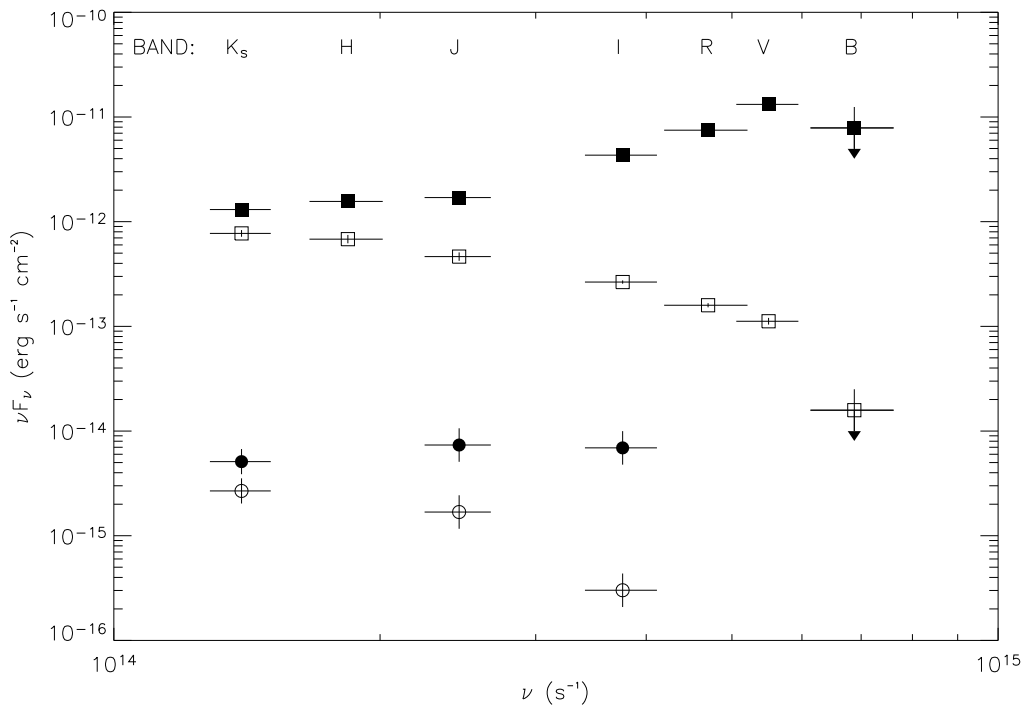}
\caption{Spectral energy distribution for 1E~0148.1$-$5937
  (circles), both as measured 
  (open) and de-reddened with $A_V=5.8$ (filled). Also
  shown is the spectral energy distribution for the brighter AXP 4U~0142+61 (squares),
  multiplied by a factor of 100 for clarity and de-redenned with
  $A_V=5.1$ (filled squares). Only detections are shown; notice that
  the B-band point for 4U~0142+61 is a marginal detection. The data
  for 4U~0142+61 is taken from Hulleman, van Kerkwijk \&
  Kulkarni (2004) and Israel et al.(2004).}\label{sed2}}
\end{center} 
\end{figure}

Figure \ref{sed2} shows the inferred spectral energy distribution (SED),
de-redenned using $A_V=5.8$ (the value derived from the X-ray
spectral fit). As
can be seen, a line of $\nu F_{\nu} \simeq const$ would be
consistent with these points (the H-and VR-band limits are consistent
with such a hypothesis).

Also on Figure \ref{sed2} is plotted the SED of 4U~0142+61 (Hulleman,
van Kerkwijk \& Kulkarni, 2004), showing the
optical and infrared regions for comparison, de-redenned with the
nominal value for that source of $A_V=5.1$. The inferred SEDs of the two AXPs
are clearly inconsistent (see particularly the I-band point).

It is worth noting here, following from Section \ref{red}, that the
reddening to both these objects is rather uncertain.  With $A_V=5.8$,
1E~1048.1$-$5937 has as SED with $\nu F_\nu=\nu^\alpha$, $\alpha =
0.2(4)$. This changes to $\alpha = -1.3(4)$ for $A_V\sim2$. Hulleman, van Kerkwijk \&
Kulkarni (2004) suggest that for 4U~0142+61, values of reddening as low
as $A_V\simeq2$ are possible, based on the run of reddening with
distance along the line of sight.  Were one to take $A_V\simeq2$ for
4U~0142+61, its SED would look much flatter and not unlike that which
is shown for 1E~1048.1$-$5937 with $A_V=5.8$. Thus it could indeed be
that the two have similar intrinsic spectra but that the value assumed
for reddening is wrong for either or both of the sources.

If we were to assume that the intrinsic spectra of the two AXPs
1E~1048.1$-$5937 and 4U~0142+61 are the same, then this allows us to
calculate the relative reddening between the two objects, $\Delta
A_V$. This value enables a comparison between the two objects without
knowledge of the specific reddening to either one. By comparing the
$I-K$ colours of the two objects we find $\Delta E(I-K) = (I-K)_{1048}
- (I-K)_{0142}=$1.2(5), implying $\Delta A_V = 2.5(5) $. Thus, if the
intrinsic spectra are indeed the same, 1E~1048.1$-$5937 would have
$V=28.2(5)$ and $R=26.9$. These predictions can be tested.

\subsection{X-ray to Infrared Flux Ratio}
In Table \ref{Fx-Fir} we list the currently known X-ray to infrared
flux ratios of the AXPs, corrected for extinction using the nominal
values of $N_H$ derived from fits to the X-ray spectra. Note that
neither the 2\ldots10keV range nor the K-band suffer considerable
extinction, so these numbers are not heavily dependent on the values
of reddening chosen. However, some of the sources have shown large
variability in both their X-ray and infrared fluxes, and the numbers
here represent values only at some particular time.

Given the variability, the flux ratios are remarkably consistent,
except for that of 1RXS~J170849$-$400910. Interestingly, the
counterpart for the latter is the least secure amongst the AXPs, as
its association is based on astrometry and possible peculiar colours
only (Israel et al., 2003). In fact, its inconsistency with the other
sources in this table suggests that perhaps the other, fainter source
within the {\em Chandra} error circle (named Star `B' by these
authors) could be the counterpart. The relative contributions of the
sources to the J-band flux is unknown, and might account for the
apparent strange colours of Star A.

1E 1048.1$-$5937 has varied in both
its X-ray and infrared emission (see Section \ref{varysec}) by over a factor of 2, so
the range which the flux ratio
could take, when measured at some particular epoch is large. From
Figure \ref{vary}, we see that the ratio of pulsed X-ray flux to
infrared flux has varied from $8\times10^{12}$\ldots$1\times10^{14}$
(in units of cts\,PCU$^{-1}$\,/\,erg\,cm$^{-2}$) . On the
one hand, then, the X-ray to
infrared flux ratios of AXPs are similar, and for 1E~2259+586 the X-ray
and infrared fluxes were correlated following an outburst. On the
other hand, the infrared and X-ray fluxes of 1E~1048.1$-$5937 seem,
if anything, to be anti-correlated. These two statements appear hard to reconcile
with one-another.

\begin{deluxetable}{lccccc}
\tablecaption{Comparison of the un-absorbed X-ray flux
  and infrared flux for all
  Anomalous X-ray Pulsars with infrared counterparts. \label{Fx-Fir}}
\tablewidth{0pt}
\tablehead{
\colhead{AXP} & \colhead{X-ray flux\tablenotemark{a}} & \colhead{$K$\tablenotemark{b}}
  & \colhead{$N_H$\tablenotemark{c}} &
  \colhead{K-band flux\tablenotemark{d}} & \colhead{$F_X/F_{K}$} \\
& \colhead{(erg s$^{-1}$ cm$^{-2}$)}& &($10^{22}$ cm$^{-2}$) & \colhead{(erg s$^{-1}$  cm$^{-2}$)} & \colhead{/1000}}
\startdata
1E 1048.1$-$5937 & $1.4 \times 10^{-11}$ & 21.3 &1.0& $5.1 \times 10^{-15}$ & 2.7 \\
4U 0142+61 & $8.3 \times 10^{-11}$ &20.1&0.91& $1.4 \times 10^{-14}$ & 5.9 \\
1RXS J170849$-$400910 & $6.4 \times 10^{-11}$&18.3&1.4 & $9.7 \times
  10^{-14}$ & 0.66 \\
1E 2259+586 & $2.0 \times 10^{-11}$ &21.7&1.1& $3.5 \times 10^{-15}$ & 5.7 \\
XTE J1810$-$197 & $2.2 \times 10^{-11}$ &20.8&1.1& $8.1 \times 10^{-15}$ & 2.7 \\
\enddata
\tablecomments{Data from Woods \& Thompson (2004) and references
  therein. The  values are not thought
  to be within active periods.}
\tablenotetext{a}{Un-absorbed flux in the 2-10keV range.}
\tablenotetext{b}{Apparent $K_\textrm{s}$ magnitude.}
\tablenotetext{c}{Inferred from fits to the X-ray spectrum.}
\tablenotetext{d}{$\nu F_\nu$, de-redenned assuming $A_V=N_H/1.79 \times
  10^{21}$ (Predehl \& Schmitt, 1995).}
\end{deluxetable}

\section{Discussion}
In this work, we have detected the likely quiescent counterpart to the
AXP 1E~1048.1$-$5937. To summarise our findings, we have found large
($\sim$2mag) variations in the infrared
brightness of 1E~1048.1$-$5937 which suggest a possible anti-correlation with
X-ray flux, but no clear relationship with spin-down rate. 
This variability is the largest that
has been seen in the AXPs, but variability is not unique to this
source. The anti-correlation between infrared
and X-ray flux is opposite to the behaviour of 1E~2259+586, where a
positive correlation was found (Tam et al., 2004).

The infrared spectrum has remained of a consistent shape with WC02. We
find that the inferred spectral energy distribution of
1E~1048.1$-$5937 is inconsistent with that inferred for 4U~0142+61,
but that they can be made consistent with an appropriate choice of
relative reddening (for which there may be justification). We find
that although both the infrared and X-ray for this source vary a large
amount and are not correlated, that the X-ray to infrared flux ratio
is consistent with the other AXPs. Perhaps this challenges our
understanding of whether the AXPs do indeed have quiescent periods,
and of the time-scales on which the magnetosphere reacts to internal
magnetic changes. We now discuss what relationships might be expected
from the magnetar model.

The model proposed for how infrared and optical emission might arise
from a magnetar is as follows. The decay of the internal magnetic
field of the magnetar induces a twist in the magnetosphere of the
magnetar. This large-scale twist is supported by global, persistent
currents, which exist out to large radii. These currents power the
non-thermal X-ray and gamma-ray luminosity, and may also produce
infrared and optical emission, through ion cyclotron
radiation at large radii (Thompson, Lyutikov \& Kulkarni, 2002).

In the case that the infrared radiation is indeed cyclotron emission by ions
gyrating in the outer magnetosphere, the infrared flux and torque
should be correlated because of the following argument. First, the
number of ions travelling to large enough radii to have Landau
transition energies in the optical/infrared range is a fixed fraction
of the total electric current. The total current in the
magnetosphere defines the amount of twist in the global magnetic field
structure (that is, the departure from a pure dipolar field), which in
turn affects the rate at which the field strength falls off with
radius. Since the torque is caused by the field strength at
the light cylinder, it must then be correlated to the ion population
at large radii and so to the infrared flux (Thompson, Lyutikov \&
Kulkarni, 2002).

Whether the infrared and X-ray flux are expected to be correlated
depends on the details of their respective emission
mechanisms. Changes in X-ray flux would be possible, for example, from
internal heating, which would not directly affect the infrared emission.

In order to constrain the emission mechanism for the infrared flux, it
will be important to establish
whether the anti-correlation between infrared and X-ray flux is indeed
real, and whether the IR flux to torque
correlation can really be ruled out. It would also be interesting to
look for a correlation with the flux in the strong, hard X-ray tail of
the spectrum that was recently discovered by Kuiper at
al. (2004). Furthermore, it would be worthwhile to seek further
spectral signatures, and to find or at least constrain the reddening
and distance to the various sources.

\medskip\noindent{\bf Acknowledgements:} MD would like to
thank Cees Bassa for help with producing astrometric figures, Fotis
Gavriil for providing XTE results, Chris Thompson for discussion, the
VizieR Service (provided by CDS and ESA) and
Victor at Las Campanas. This work is supported by NSERC.

\end{document}